\documentclass[11pt,a4paper]{article}

\usepackage{slashed}
\usepackage{amsmath}
\usepackage[pdfstartview=FitH,
colorlinks=true,linkcolor=red,anchorcolor=black,citecolor=blue,urlcolor=black,filecolor=blue]{hyperref}
\usepackage[english]{babel}
\usepackage[top=2.3cm,right=2.3cm,left=2.3cm,bottom=2.3cm]{geometry}
\usepackage{amsmath,amssymb,titling,authblk}
\usepackage{slashed}
\usepackage{amsmath}
\usepackage{amscd}
\usepackage[normalem]{ulem}
\usepackage{appendix}
\usepackage{bbold}
\usepackage{cite}

\usepackage{afterpage}
\usepackage{cite}

\allowdisplaybreaks[4]
\usepackage{color}  

\definecolor{verde}{cmyk}{.83,.21,1,.08}
\definecolor{darkorchid}{rgb}{0.6, 0.2, 0.8}

\newcommand{\be}{\begin{equation}}
\newcommand{\ee}{\end{equation}}
\newcommand{\bea}{\begin{eqnarray}}
\newcommand{\eea}{\end{eqnarray}}
\newcommand{\ii}{\mathrm{i}}

\newcommand{\dd}{\mathrm{d}}
\newcommand{\del}{\partial}
\newcommand{\la}{\label}

\newcommand{\R}{\mathbb{R}}

\newcommand{\Z}{\mathbb{Z}}
\newcommand{\eqn}[1]{(\ref{#1})}

\numberwithin{equation}{section}

\title{Four-dimensional noncommutative deformations of $U(1)$ gauge theory and $L_{\infty}$ bootstrap.}
\author[1,2]{Maxim Kurkov}
\author[1,2]{Patrizia Vitale}
\affil[ ]{}
\affil[1]{\textit{\footnotesize Dipartimento di Fisica ``E. Pancini'', Universit\`a di Napoli Federico II, Complesso Universitario di Monte S. Angelo Edificio 6, via Cintia, 80126 Napoli, Italy.}}
\affil[2]{\textit{\footnotesize INFN-Sezione di Napoli, Complesso Universitario di Monte S. Angelo Edificio 6, via Cintia, 80126 Napoli, Italy.}}
\affil[ ]{}
\affil[ ]{\footnotesize e-mail: \texttt{ max.kurkov@gmail.com, patrizia.vitale@na.infn.it}}

\begin{document}
\maketitle
\begin{abstract}
We construct a family of four-dimensional noncommutative deformations of $U(1)$ gauge theory following a general scheme, recently proposed in JHEP 08 (2020) 041 for a class of coordinate-dependent noncommutative  {algebras}. This class includes  the $\mathfrak{su}(2)$,  the  $\mathfrak{su}(1,1)$ and the angular (or $\lambda$-Minkowski) noncommutative structures. We find that the  presence of a fourth, commutative coordinate $x^0$ leads to substantial novelties in the expression for the deformed field strength with respect to the corresponding three-dimensional case.  {The constructed field theoretical models are Poisson gauge theories, which correspond to the semi-classical  limit of fully noncommutative gauge theories.}    Our expressions for the deformed gauge transformations,  the deformed field strength and the deformed classical action  exhibit  flat commutative limits and they are exact in the sense that all orders in the deformation parameter are present. We review the connection of the formalism with the $L_{\infty}$ bootstrap and with  symplectic embeddings, and derive the $L_{\infty}$-algebra, which underlies our model.
\end{abstract}

\section{Introduction}
Noncommutative gauge and field theories have been widely studied over more than twenty years. Much has been written about physical motivations for considering space-time to be ``quantum'' and physical models to be described in terms of noncommuting observables, if one wants to go beyond the dichotomy between classical gravity  and quantum physics. There exist excellent reviews for that, 
 {see for instance~\cite{Szabo:2001kg,Aschieri:2009zz,Wulkenhaar:2019tlq} 
}. 
Despite the large efforts made, there is however  no general consensus about the  appropriate noncommutative generalisation of field theory, mainly because, except for very few models, all attempts proposed present formal and/or interpretative problems, which render  the results   not fully satisfactory.  Nonetheless, the problems addressed with the promise of providing effective models of space-time quantisation and compatible  gauge theories  maintain their validity. 

One main motivation for confronting once again with noncommutative gauge theory is a series of recent publications proposing the framework of $L_\infty$ algebras and a bootstrap approach as appropriate for formulating gauge noncommutativity in a consistent way   \cite{Blumenhagen:2018kwq, Kupriyanov:2019ezf}.  Moreover, an interesting connection has been established between the $L_\infty$ bootstrap and symplectic embeddings of non-commutative algebras \cite{Kupriyanov:2021cws, Kupriyanov:2021aet}. It is worth  mentioning  that the role of $L_\infty$ algebras in gauge and field theory is already investigated in \cite{Stasheff2002, Hohm:2017pnh, Jurco:2018sby}, see also \cite{Ciric:2021rhi,Grewcoe:2020hyo,Ciric:2020hfj, Arvanitakis:2019ald, Blumenhagen:2018hvs,Hohm:2017cey} for recent progresses in studies of the $L_{\infty}$ structures in the field theoretic context.

In the present paper, we shall take advantage of the   constructive   approach proposed  in \cite{Kupriyanov:2020sgx} which, starting from the request that  gauge theory be compatible with the desired space-time noncommutativity {\it and} be equivalent to  the standard one in the commutative limit, yields recursive equations for field dependent gauge transformations and deformed field strength.  As we shall explicitly discuss, the procedure {is closely related to}
 the $L_\infty$ bootstrap and symplectic embedding approaches.

{The only exact (all orders in the non-commutativity parameter) nontrivial\footnote{i.e. with coordinate-dependent noncommutativity} models, which have been constructed so far along the lines of~\cite{Kupriyanov:2020sgx}, and exhibiting the flat commutative limit, are the three-dimensional $U(1)$ theory with  $\mathfrak{su}(2)$ noncommutativity~\cite{Kupriyanov:2020sgx} and the two-dimensional $U(1)$ model with  kappa-Minkowski\footnote{The deformed gauge transformations and the deformed field strength for the kappa-Minkowski noncommutativity  have been constructed in~\cite{Kupriyanov:2020axe} in arbitrary dimension $d$, however, the  gauge invariant classical action exhibits the \emph{flat} commutative limit at $d=2$ only.  } noncommutativity~\cite{Kupriyanov:2020axe}. Therefore, the construction of four-dimensional models of this kind seems to be a valuable and timely problem.}

{In the present work we fill this gap:} we construct exact noncommutative four-dimensional deformations of $U(1)$ gauge theory, implementing several three-dimensional  {noncommutative} structures within the general framework proposed in~\cite{Kupriyanov:2020sgx},  {and adding one more commutative coordinate. As we shall see below, such an addition brings somewhat more than a naive generalisation of the corresponding three-dimensional setup.} For the non-trivial sector of the algebra we shall consider explicitly the angular (or $\lambda$-Minkowski) noncommutativity~\cite{Angular0,Angular1,Angular2,Angular3,Angular4,Angular6,Angular5},  
\be
[\hat{x}^3,\hat{x}^1]  = -\ii \,  \lambda \hat{x}^2, \quad [\hat{x}^3,\hat{x}^2] = \ii \,  \lambda \hat{x}^1, \quad [\hat{x}^1,\hat{x}^2]  = 0, \la{aNC}
\ee
and the $\mathfrak{su}(2)$ noncommutativity~\cite{Hammou2002,  selene2002, Galikova:2013zca, Kupriyanov:2012nb, Vitale:2012dz, Gere:2013uaa, Vitale:2014hca, Kupriyanov:2015uxa, Gere:2015ota, Juric:2016cfp},
\be
[\hat{x}^{k}, \hat{x}^{l}] = - \ii \,  \lambda \,\varepsilon^{kls}\,\delta_{sp} \,\hat x^{p}. \la{su2NC}
\ee
 The latter may be easily generalised to $\mathfrak{su}(1,1)$, while the  time variable $\hat{x}^0$ stays commutative for all cases considered. We shall use the Greek letters $\mu$, $\nu$, ..., and the Latin letters $a$, $b$, $c$, ...,  to denote the four-dimensional and the three-dimensional (i.e. the spatial) coordinates respectively. The three-dimensional deformation\footnote{Our notations do not coincide with the ones of~\cite{Kupriyanov:2020sgx}:  $\lambda_{\mathrm{our}} = -2\theta_{\mathrm{Ref.}\mbox{\cite{Kupriyanov:2020sgx}}}  $. } of $U(1)$ gauge theory, based on the $\mathfrak{su}(2)$ noncommutativity~\eqref{su2NC} has already been studied in detail in~\cite{Kupriyanov:2020sgx}. We shall see, however, that an addition of time as a  fourth commutative coordinate extends the results of~\cite{Kupriyanov:2020sgx} in a nontrivial way.  
{For a given starting space-time  $\mathcal{M}$}, we shall indicate   with  $\mathcal{A}_\Theta=(\mathcal{F}(\mathcal{M}),\star)$  the noncommutative algebra of functions representing noncommutative space-time,  {equipped by some noncommutative star product
\be
f\star g \neq g\star f, \quad\quad f,g\in \mathcal{A}_\Theta
\ee
which, for coordinate functions, reproduces the linear algebras \eqn{aNC} and \eqn{su2NC}.
  Noncommutativity is therefore specified by  the $x-$dependent skew-symmetric  matrix  $\Theta(x)$:  
 \be
[{x}^{\mu}, x^{\nu}]_{\star}  = \ii\,  \Theta^{\mu\nu}({x}),
\ee
which we assume %that $\Theta^{\mu\nu}$ 
to be  a Poisson bivector in order to maintain  associativity of the star-product.  {The symbol $[\,\, ,\,\, ]_{\star}$ denotes the star commutator, defined as follows
\be
[f,g]_{\star} \equiv f\star g - g\star f,\quad\quad \forall f,g\in  \mathcal{A}_\Theta. 
\ee
}
Standard   $U(1)$ gauge transformations,
$
\delta^0_f A=\partial f, 
$\,
with $f\in \mathcal{F}(\mathcal{M})$,  close an  Abelian algebra, $[\delta^0_f,\delta^0_g] = 0$. For non-Abelian gauge theories where  gauge parameters are  valued in a non-Abelian Lie algebra , ${\mathbf f} = f_i \tau^i$, we have instead $\delta^0_{\mathbf f} A= \del {\mathbf f} -i  [A, {\mathbf f}] $ so that 
$$[\delta^0_{\mathbf f},\delta^0_{\mathbf g}] A= \del [{\mathbf f}, {\mathbf g}] -i  [A,[{\mathbf f},{\mathbf g} ]]= \delta^0_{ [{\mathbf f}, {\mathbf g}]}A.$$ 
Namely, the algebra of gauge transformations closes with respect to a non-Abelian Lie bracket. Noncommutative  $U(1)$ gauge theory, with gauge parameters now  belonging to $\mathcal{A}_\Theta$ behaves very much like non-Abelian  theories. Therefore, according to   \cite{Kupriyanov:2020sgx},
we shall require that the algebra of gauge transformations closes with respect to the star commutator, namely 
\begin{equation}
[\delta_f,\delta_g]A=\delta_{-\ii[f,g]_\star}A. \la{gaugealgebra0}
\end{equation}
However, if gauge  transformations are defined as a natural generalisation of the non-Abelian case,
\be\label{ncgautr}
A'= A+ \del f -\ii [A, f]_\star,
\ee
it is known that, by composing two such transformations, we get the result \eqn{gaugealgebra0} only if $\del$ is a derivation of the star commutator, which in general is not the case. }
Hence, the guiding principle in~\cite{Kupriyanov:2020sgx} was the definition of the infinitesimal gauge transformations,
\be
A_{\mu} \longrightarrow A_{\mu} + \delta_f A_{\mu},
\ee
in such a way that they  close the {noncommutative} algebra \eqn{gaugealgebra0}
and  reduce to the standard $U(1)$ transformations in the commutative limit, 
\be
\lim_{\lambda\rightarrow 0}\delta_f A_{\mu} = \partial_{\mu}f . \la{comlim}
\ee

%Throughout this paper we work at the semi-classical limit. 
The star commutator, which enters in~\eqref{gaugealgebra0} has the following structure: 
\be
[f,g]_{\star} = \ii\,   \{f,g\} + ...,
\ee
where $\{f,g\}$ stands for the  Poisson bracket of $f$ and $g$,
\be
\{f,g\} \equiv \Theta^{\mu\nu}\, \partial_{\mu}f \,\partial_{\nu}g  , \la{PoissBrack}
\ee 
while the remaining terms, denoted through ``...", contain higher derivatives.  {From now on we neglect  these terms, namely we  consider  the semi-classical limit~\cite{Kupriyanov:2021cws,Kupriyanov:2021aet}. } 
 Therefore our noncommutative gauge algebra becomes  {the Poisson gauge algebra}:
\be
[\delta_f, \delta_g] = \delta_{\{f,g\} }. \la{gaugealgebra}
\ee

%%%%%%%%%%%%%%%%%%%%%
%%%%%%%%%%%%%%%%%%%%%
 We shall consider in what follows a  {two-parameter} family of  Poisson structures:
 {\be
\Theta^{0\mu} = 0 = \Theta^{\mu 0}, \quad \Theta^{jk} =- \lambda\, \varepsilon^{jks}\,\check\alpha_{s l} \,x^l, \la{Ps}
\ee}
where the $3\times 3$ matrix $\check{\alpha}$ is defined as follows:
 {\be
\check{\alpha} := \mathrm{diag}\,\{1,\,1,\,\alpha \}, \quad \alpha\in\mathbb{R}.
\ee}
At $\alpha = 0$ we get the Poisson structure which corresponds to the angular noncommutativity, 
\be
\{{x}^3,{x}^1\} = - \lambda {x}^2, \quad \{{x}^3,{x}^2\} = \lambda {x}^1, \quad \{{x}^1,{x}^2\} = 0, \quad \{x^j, x^0\} = 0,
\ee
while  at $\alpha = 1$ the three-dimensional bivector $\Theta^{jk}$ is nothing but the Poisson structure of the $\mathfrak{su}(2)$ case,
\be
\{{x}^{k}, {x}^{l}\} = -  \lambda \,\varepsilon^{kls}\,\delta_{sp} \, x^{p}. \ee
Another interesting case is represented  by $\alpha= -1$ which corresponds to the Lie algebra $\mathfrak{su}(1,1)$.
We emphasise however that the Jacobi identity,
\be
f^{\xi\lambda}_{\mu} f^{\nu\phi}_{\lambda}  + f^{\nu \lambda}_{\mu} f^{\phi\xi}_{\lambda} + f^{\phi\lambda}_{\mu}f^{\xi\nu}_{\lambda} = 0, 
\quad\quad f^{\xi\lambda}_{\mu} \equiv \partial_{\mu}\Theta^{\xi\lambda},
\ee
is satisfied for  \emph{any} $\alpha$,  not just at $\alpha = 0$ and $\alpha=\pm 1$. Introducing 
 the projector $\boldsymbol{\delta}_{\mu}^{\nu}$ on the three-dimensional space,
 {\be
\boldsymbol{\delta}_{\mu}^{\nu} := \delta_{\mu}^{\nu} - \delta_{\mu}^0\delta^{\nu}_0,
\ee}
we get an explicit formula for the structure constants: 
 {\be
 f^{\mu \nu}_{\rho}  = - \lambda\, \boldsymbol{\delta}^{\mu}_{j}\,\boldsymbol{\delta}^{\nu}_{k}\,\boldsymbol{\delta}^{l}_{\rho}\,\varepsilon^{j k s}\,\check\alpha_{s l}, \la{ourf}
\ee} 
 {yielding
\be
\{x^j, x^k \} = - \lambda \,\varepsilon^{j k s}\,\check\alpha_{s l}\, x^s, \quad\quad \{x^j, x^0 \} = 0.
\ee
}
The paper  is organised as follows.   Sec.~\ref{NCgauge}  is devoted to deformed gauge transformations. Moreover, a connection with the symplectic embedding approach is discussed.  In Sec. \ref{Linf} we present   {relevant aspects} of the $L_{\infty}$ bootstrap approach to  gauge theories  {and establish the $L_{\infty}$ algebra, which corresponds to our gauge transformations.} In Sec.~\ref{Fsec} we introduce a deformed field strength and a suitable classical action.

\section{Noncommutative gauge transformations.}\label{NCgauge}
According to~\cite{Kupriyanov:2020sgx} the infinitesimal deformed gauge transformations, which close the algebra~\eqref{gaugealgebra}, and reproduce the correct undeformed limit~\eqref{comlim},  can be constructed by allowing for  a field-dependent deformation as follows:
\be
\delta_f A_{\mu} = \gamma_{\mu}^{\nu}(A)\partial_\nu f + \{A_{\mu},f\}. \la{gauge0}
\ee
 {This variation satisfies the following derivation property~\cite{Kupriyanov:2021cws}:
\be
\delta_{fg} A_{\mu} = g\delta_f A_{\mu} + f\delta_g A_{\mu}.
\ee
}
For \eqn{gaugealgebra} to be satisfied  the $4\times 4$ matrix  $\gamma$ has to solve the master equation\footnote{We use the notation $\partial^{\mu}_A \equiv \frac{\partial}{\partial A_{\mu}}$.}\label{footnote2}
\be
\gamma_{\mu}^{\nu} \partial^{\mu}_A \gamma^{\xi}_{\lambda} - \gamma^{\xi}_{\mu} \partial_A^{\mu} \gamma^{\nu}_{\lambda} 
+ \Theta^{\nu\mu} \partial_{\mu} \gamma^{\xi}_{\lambda} 
- \Theta^{\xi\mu} \partial_{\mu} \gamma^{\nu}_{\lambda} 
- \gamma^{\mu}_{\lambda}\partial_{\mu}\Theta^{\nu\xi} = 0, \la{master1}
\ee
moreover, it has to reduce to the identity at the commutative limit,
\be
\lim_{\lambda\rightarrow 0} \gamma_{\nu}^{\mu} = \delta_{\nu}^{\mu}. \la{comlim1}
\ee
The last requirement guarantees that the noncommutative transformations~\eqref{gauge0} reproduce  the standard Abelian gauge transformations~\eqref{comlim} in the undeformed theory.
A general result has been established in  %~(6.3) of~
\cite{Kupriyanov:2021cws} in the context of symplectic embeddings, that  is  valid for any $\Theta$ which is linear in $x$.
 It  suggests a solution of Eq.~\eqref{master1} in the form\footnote{ Note that our notations differ from the ones of~\cite{Kupriyanov:2021cws}. In order to obtain our Eq.~\eqref{gensolu} one has to set $t=1$, $p=A$  and replace $\gamma$ by $\gamma  - \mathbb{1}$ in Eq.~(6.3) of~\cite{Kupriyanov:2021cws}. 
 }\label{footnote3}
\be
\gamma_{\mu}^{\nu}(A) = -\frac{1}{2}f_{\mu}^{\nu\lambda}A_{\lambda} + \delta_{\mu}^{\nu} + \chi\left(-\frac{M}{2}\right)_{\mu}^{\nu}, \la{gensolu}
\ee
where the matrix $M$ is defined by 
\be
M_{\mu}^{\nu} = f_{\xi}^{\nu\lambda}f_{\mu}^{\xi \phi} A_{\lambda}A_{\phi} \la{Mdef}
\ee
and the function $\chi$ reads
\be
\chi(u) = \sqrt{\frac{u}{2}} \cot{\sqrt{\frac{u}{2}}} - 1  = \sum_{n=1}^{\infty}\frac{(-2)^n \,B_{2n} \,u^n}{(2n)!}.
\ee
In the last equalities the quantities $B_{2n}$ are the Bernoulli numbers, and $\chi\left(\frac{M}{2}\right)_{\mu}^{\nu}$ have to be understood as the matrix elements of $\chi(M/2)$. 

By replacing the structure constants ~\eqref{ourf} in the definition~\eqref{Mdef} we obtain
\be
 M_{\mu}^{\nu}  = -\lambda^2\, {  \mathrm{\boldsymbol{Z}} }\,\hat{M}_{\mu}^{\nu},
\ee
where the matrix $\hat{M}$, defined by 
\be
\hat{M}_{\mu}^{\nu} =  {\delta}^{\nu}_{\mu} - \frac{\boldsymbol{A}_{\boldsymbol{\alpha}}^{\nu} \boldsymbol{A}_{\mu}}{ {  \mathrm{\boldsymbol{Z}} }},
\ee
is a \emph{projector}, i.e. $\hat{M}^2 = \hat{M}$,
and hence
\be
\hat{M}^n = \hat{M}, \quad\quad \forall n \in\mathbb{N}. \la{projprop}
\ee
From now on we shall make use of the  notations\footnote{Do not confuse $\hat{\alpha}$ and $\check{\alpha}$! } :
 {\bea
\boldsymbol{A}_{\mu}  &:=& \boldsymbol{\delta}_{\mu}^{\nu} A_{\nu}, \nonumber\\
\boldsymbol{A}_{\boldsymbol{\alpha}}^{\mu} &:=& \hat{\alpha}^{\mu \xi} A_{\xi} , \quad
\hat{\alpha} := \mathrm{diag}\{ \alpha,\,\alpha,\, 1,\,0\}, \nonumber\\
 {  \mathrm{\boldsymbol{Z}} } &:=& A_{\mu} \boldsymbol{A}_{\boldsymbol{\alpha}}^{\mu} =  \alpha\cdot \left(A_1\right)^2 + \alpha\cdot \left(A_2\right)^2 +  \left(A_3\right)^2. \la{Znotations}
\eea}
Using the identity~\eqref{projprop} we can easily calculate the nontrivial term of Eq.~\eqref{gensolu}:
\bea
\chi\left(-\frac{M}{2}\right) &=& \sum_{n=1}^{\infty}\frac{(-2)^n \,B_{2n} }{(2n)!} \left(-\frac{M}{2}\right)^n 
= \sum_{n=1}^{\infty}\frac{(-2)^n \,B_{2n} }{(2n)!} \left( \frac{\lambda^2\, {  \mathrm{\boldsymbol{Z}} }\,}{2}\right)^n \cdot\underbrace{\hat{M}^n}_{\hat{M}} \nonumber \\
&=& \left(\sum_{n=1}^{\infty}\frac{(-2)^n \,B_{2n} }{(2n)!} \left( \frac{\lambda^2\, {  \mathrm{\boldsymbol{Z}} }\,}{2}\right)^n \right) \cdot{\hat{M}}  = \chi\left(\frac{\lambda^2\, {  \mathrm{\boldsymbol{Z}} }\,}{2}\right) \cdot{\hat{M}}.
\eea
Substituting this result in the general formula~\eqref{gensolu}, we arrive at  the final expression for $\gamma$, 
 {\bea
\gamma^{\nu}_{\mu}(A) &=&- \frac{1}{2}\, f^{\nu\lambda}_{\mu}\,   A_{\lambda} + \delta^{\nu}_{\mu} 
+ \frac{\lambda^2}{4}\, \hat{\chi}\left(\frac{ {  \mathrm{\boldsymbol{Z}} }\,\lambda^2}{4}\right) \left( {  \mathrm{\boldsymbol{Z}} }\,\boldsymbol{\delta}^{\nu}_{\mu} - \boldsymbol{A}_{\boldsymbol{\alpha}}^{\nu} \boldsymbol{A}_{\mu} \right), \la{GammAsol}
\eea}
where we introduced another form factor,
 {\bea
\hat{\chi}(v) &:=& \frac{1}{v} \big( \sqrt{v}\, \mathrm{cot}\sqrt{v}    -1\big), \la{hchidef}
\eea}
in order to confront our results with the ones of~\cite{Kupriyanov:2020sgx}. Setting $\alpha = 1$, one can easily see that the three-dimensional part of~\eqref{GammAsol}, viz  $\gamma_{j}^i$, coincides with the known three-dimensional result~(2.11) of~\cite{Kupriyanov:2020sgx} for the $\mathfrak{su}(2)$-case.

Interestingly, the field-dependent deformation of gauge transformations (\ref{gauge0}) has been derived in \cite{Kupriyanov:2021cws} as the result of a symplectic embedding of the Poisson manifold $(\mathcal{M}, \Theta)$ into the symplectic manifold $(T^*\mathcal{M}, \omega)$, where $T^*\mathcal{M}$ denotes the cotangent bundle and $\omega$ an appropriate symplectic form such that $\pi_* \omega^{-1}= \Theta$, $\pi:T^*\mathcal{M}\rightarrow \mathcal{M}$ being the projection map.  

Shortly, the idea of symplectic embeddings of Poisson manifolds is a generalization of  symplectic realizations \cite{weinstein,cranic}, which   consist in the following. One considers the canonical symplectic form $\omega_0$ on $T^*\mathcal{M}$, which is locally given by  {$\omega_0= d\lambda_0= dp_{\mu} \wedge dx^{\mu}$ }with  {$\lambda_0=p_{\mu} d x^{\mu}$} the Liouville one-form. The contraction of $\lambda_0$ with the Poisson tensor $\Theta$ defined on $\mathcal{M}$ yields a vector field,  {$X^\Theta= \Theta(\lambda_0,\;)= \Theta^{\mu\nu}(x) p_{\mu} \del_{x^{\nu}}$} whose flow we shall indicate with $\varphi^\Theta_t, t\in\R$. In terms of the latter, it is possible to endow, at least locally, the cotangent space  with a new symplectic form, $\omega$ whose inverse naturally projects down to the Poisson tensor $\Theta$ on $\mathcal{M}$ through the projection map $\pi:T^*\mathcal{M}\rightarrow \mathcal{M}$. According to  \cite{weinstein,cranic} such a form is given by the integrated pull-back of the canonical symplectic form $\omega_0$ through the flow associated with the vector field $X^\Theta$,
\be\label{omegadef}
\omega := \int_0^1 (\varphi^\Theta_t)^*(\omega_0) \, dt
\ee
which in coordinates reads $\omega= dy^\mu \wedge dp_\mu $ with  {$y^\mu(x,p)= \int_0^1 x^\mu \circ \varphi^\Theta_t \, dt$}. The Jacobian matrix $J= (\del y^\mu/\del x^\nu)$ is formally invertible. On denoting its inverse  by  $\gamma(x,p)$  the symplectic Poisson tensor $\omega^{-1}= \frac{\del}{\del y^\mu} \wedge \frac{\del}{\del p_\mu}$ is given in terms of the original variables $(x,p)$ and the   matrix $\gamma(x,p)$, according to 
\be\label{poitens}
\omega^{-1}=  \Theta^{\mu\nu}\frac{\del}{\del x^\mu} \wedge\frac{\del}{\del x^\nu}-\gamma^\mu_\nu(x,p) \frac{\del}{\del x^\mu} \wedge\frac{\del}{\del p_\nu}.
\ee
Let's pose $\omega^{-1}= \Lambda$. A generalization of the previous procedure consists in {\it defining} $\Lambda$ as a deformation of $\Theta$, according to Eq. \eqn{poitens} and imposing that it satisfies Jacobi identity, provided $\Theta$ does. This amounts to compute  
 the Schouten bracket, $[\Lambda,\Lambda]$,  and impose  that it be zero. We obtain  the following  equation for the matrix $\gamma$, 
\be
\gamma_{\mu}^{\nu} \frac{\del}{\del p_\mu} \gamma^{\xi}_{\lambda} - \gamma^{\xi}_{\mu} \frac{\del}{\del p_\mu}  \gamma^{\nu}_{\lambda} 
+ \Theta^{\nu\mu} \frac{\del}{\del x^\mu}  \gamma^{\xi}_{\lambda} 
- \Theta^{\xi\mu} \frac{\del}{\del x^\mu}  \gamma^{\nu}_{\lambda} 
- \gamma^{\mu}_{\lambda}\frac{\del}{\del x^\mu} \Theta^{\nu\xi} = 0, \label{masterp1}
\ee
where $[\Theta,\Theta]=0$ has been used. 
The latter is exactly the master equation \eqn{master1} after  replacing  derivatives with respect to $p_\mu$  by derivatives with respect to $A_\mu$, which is, however, a non-trivial difference, since $A_\mu$ is itself a function of $x$, while $p_\mu$ is obviously not. The relation between the two approaches, which  has been established in \cite{Kupriyanov:2021cws, Kupriyanov:2021aet}, may be summarised as follows.  
\\
Let us first consider the standard setting with $\Theta=0$. Then, the cotangent bundle $T^*\mathcal{M}$ is endowed with the canonical symplectic form $\omega_0$.  The gauge field $A\in \Omega^1(U)$,  $U\subset \mathcal{M}$  is associated with a local section $s_A: U\rightarrow T^*U$, through a local trivialisation, $\psi_{U}^{-1}(s_A(x))= (x, A(x))$.  The image of $s_A$ is a submanifold of $T^*U$. Let 
\be\label{xiA}
\xi_A= \lambda_0-\pi^*A
\ee
be a local one-form on $T^*U$ with $\lambda_0$ the Liouville form. We have  
\be s_A^*(\xi_A)=0\label{constraint}
\ee
it being $s_A^*(\lambda_0)=A= (\pi\circ s_A)^*(A)$. This means that $\xi_A$ vanishes exactly on the submanifold ${\rm im}(s_A)\subset T^*U$. Therefore the latter is identified by the constraint \eqn{constraint}, which in turn amounts to fix the fibre coordinate at $x$,  $p$  to its value $A(x)$ identified by the section $s_A$.
Then, the infinitesimal gauge transformation of the gauge potential $A$, with gauge parameter $f$, may be defined in terms of the canonical Poisson bracket $\omega_0^{-1}$ as follows
\be\label{deltacan}
\delta_f A_\nu(x) = s_A^*\{\pi^*f, \xi_{A_\nu}\}_{\omega_0^{-1}}= \frac{\del f}{\del x^\mu }\frac{\del  \xi_{A_\nu}}{\del p_\mu}= \del_\nu f.
\ee
Now let us  consider the case $\Theta\ne 0$, namely, $(\mathcal{M}, \Theta)$ is a Poisson manifold. A symplectic embedding  is performed as described above, with symplectic form now given by %Eq. \eqn{omegadef}, 
{the inverse of~\eqref{poitens},}
while  the image of $U\subset \mathcal{M}$ through the local section $s_A$ is still defined by the constraint \eqn{xiA}. Then, the infinitesimal gauge transformation of the gauge potential is formally the same as in the previous case,  \eqn{deltacan}, except for the fact that $\omega_0^{-1}$ is to be replaced by the Poisson tensor $\omega^{-1}$ defined by \eqn{poitens}.  Therefore we have
\be
\delta_f A_\mu:= s_A^*\{\pi^*f, \xi_{A_\nu}\}_{\omega^{-1}}= \Theta^{\rho\sigma} \frac{\del f}{\del x^\rho}\frac{\del \xi_{A_\mu}}{\del x^\sigma}-\gamma^\rho_\sigma(x,A) \frac{\del f}{\del x^\rho} \frac{\del \xi_{A_\mu}} {\del p_\sigma}=\{A_\mu, f\}_\Theta+ \gamma^\rho_\mu(x,A) \frac{\del f}{\del x^\rho}, 
\ee
which   is precisely the starting assumption \eqn{gauge0}. The master equation for $\gamma$ , Eq. \eqn{masterp1}, yields therefore  Eq.~\eqref{master1} once  the constraint \eqn{xiA} has been imposed.  
Notice that, in order to compare with the results   of \cite{Kupriyanov:2021cws} we have to pose  
$t=1$ and substitute $\gamma$ by $\gamma  - \mathbb{1}$.

Let us remark  that the solution of the master equation~Eq.~\eqref{master1}   is not unique, i.e. one may construct other deformed gauge transformations, which close the algebra~\eqref{gaugealgebra}. In the approach just described, this is due to the freedom in choosing different symplectic embeddings for the Poisson manifold $(\mathcal{M}, \Theta)$  \cite{Kupriyanov:2021cws, Kupriyanov:2021aet},
and  {gives rise} to a field redefinition,  {which maps gauge orbits of the original fields onto gauge orbits of the new fields~\cite{Blumenhagen:2018shf}}.

In the next section we will see that the present construction is closely related to the $L_{\infty}$-bootstrap. In that setting, the ambiguity mentioned above corresponds to a quasi-isomorphism of the underlying $L_{\infty}$ algebra, which is \emph{unique}  (up to quasi-isomorphisms)~\cite{Blumenhagen:2018shf}. 

\section{ {Relation to }$L_\infty$ algebras  and bootstrap.}\label{Linf}
$L_\infty$ algebras are  {homotopy generalisations of Lie algebras} defined on a graded vector space $V={\oplus}_k V_k, k\in \Z$, with multi-linear $n$-brackets\footnote{ {We are considering the so called $l$-picture~\cite{Hohm:2017pnh}.}}
\be
\ell_n:(v_1,\cdots,v_n)\in V^{\otimes n}\rightarrow v\in V .
\ee
 $ k\in\Z$ denotes the grading of the subspace $V_k$, so that  {${\rm deg}(v)= k \iff v\in V_k$} . 
 By definition
 \be
 {\rm deg}\big(\ell_n(v_1,\cdots,v_n)\big) = n-2 + \sum_{i=1}^{n} {\rm deg}(v_i).
 \ee
 The brackets are graded anti-symmetric
 \be
 \ell_n(v_1,\cdots, v_{j},v_{j+1},\cdots)= {(-1)^{1+{\rm deg}(v_j) {\rm deg}(v_{j+1})}}\ell_n(v_1,\cdots, v_{j+1},v_{j},\cdots) {,} \la{gas}
\ee 
 {and satisfy the generalised Jacobi identities,
\bea
\sum_{i+j = n+1} (-1)^{i(j-1)}\sum_{\sigma}(-1)^{\sigma} \epsilon(\sigma, v) \,l_j(l_i(v_{\sigma{(1)}},...,v_{\sigma{(i)}}),v_{\sigma{(i+1)}},...,v_{\sigma{(n)}}) = 0,\quad\quad n\in\mathbb{N}
\eea
where $\sum_{\sigma}$ denotes the sum over permutations, $\sigma$, of the variables $v_1,$...,$v_n$ such that, 
 {\be
{\sigma{(1)}} < \cdots < {\sigma{(i)}}, \quad\quad {\sigma{(i+1)}} < \cdots < {\sigma{(n)}},
\ee}
 { $(-1)^{\sigma}$ takes care of  the signature of the permutation
and $\epsilon(\sigma, v)$ stands for the Koszul sign, which takes into account the degree of the permuted entries  (see~\cite{Hohm:2017pnh} for details).}

 { $L_{\infty}$ 
algebras and  gauge transformations are related in the following way~\cite{Stasheff2002,Hohm:2017pnh}.} 
Consider a graded space $V$ such that the only nonempty subspaces are $V_0$ and $V_{-1}$. By construction the former is identified with a space of the gauge parameters, $f\in V_0$
whilst the latter contains the gauge fields, $A = A_{\mu}\,\dd x^{\mu} \in V_{-1}$. We shall look for the deformed gauge  transformation in the  form of a series  expansion, as follows:
\be
\delta_f A  = \sum_{n=0}^{\infty} (-1)^{\frac{n(n-2)}{2}} l_{n+1}(f,A,\cdots,A). \la{gaugegen1}
\ee
By setting 
\begin{eqnarray}
l_1(f) &=& \dd f  = (\partial_{\mu}f) \,\dd x^{\mu}, \la{linput1}\\
l_2(f,g) &=& -\left\{f,g\right\},\quad\quad \forall f,g\in V_0, \la{linput2}
\end{eqnarray}
 {and} determining the remaining brackets, $l_k$, from the requirement of  closure of the $L_{\infty}$-algebra, one can build the gauge transformation~\eqref{gaugegen1}.  {Such a ``completion" is referred to as  the  $L_{\infty}$ bootstrap~\cite{Blumenhagen:2018kwq,Kupriyanov:2019ezf}.}
General properties of the $L_{\infty}$-construction automatically insure that the condition~\eqref{gaugealgebra} is satisfied, see e.g.~\cite{Kupriyanov:2019ezf,Kupriyanov:2019cug}.

In the previous section we have constructed the deformed gauge transformations without any reference to the $L_{\infty}$ algebras, however,  Proposition 5.9 of~\cite{Kupriyanov:2021cws} guarantees, that for any  {symplectic} embedding, related\footnote{Here ``related" means that both the deformed  gauge transformation and the  {symplectic} embedding are defined via the same matrix $\gamma$, which is a solution of the master equation~\eqref{master1}. } to the deformed gauge transformation~\eqref{gauge0}, the $L_{\infty}$ algebra is indeed there, and can be constructed as follows.   

\begin{itemize}
\item{Expanding the right hand side of the transformation~\eqref{gauge0}, presented as 
\be
\delta_f A  = \big(\gamma_{\mu}^{\nu}(A)\partial_{\nu} f + \{A_{\mu},f\}\big)\,\dd x^{\mu},
\ee
in powers of $A$, and comparing  with the right-hand side of~\eqref{gaugegen1}  one finds all the brackets of the form $l_n(f,A,\cdots,A)$. All other brackets, which depend on a single argument $f$ and {$n-1$} arguments $A$ can be, obviously, recovered from the mentioned ones by the graded antisymmetry~\eqref{gas}.  
}
\item{The only nonzero bracket, which involves \emph{two} arguments $f,g\in V_0$, is given by Eq.~\eqref{linput2}. }
\item{All other brackets are identically equal to zero.} 
\end{itemize}
The proposition, mentioned above,  also asserts that   $L_{\infty}$-algebras which correspond to different choices of $\gamma$ (i.e. different  {symplectic} embeddings), associated with the same Poisson bivector $\Theta$ via Eq.~\eqref{master1}, are necessarily connected by $L_{\infty}$-quasi-isomorphisms. From this point of view the $L_{\infty}$ structure, which underlies a given deformed gauge  transformation~ {of the form \eqref{gauge0}} is ``unique".

Applying the prescription presented above to the matrix $\gamma$, given by Eq.~\eqref{GammAsol}, we get 
\bea
\delta_f A  &=& (\partial_{\mu}f)\dd x^{\mu} +  \left\{A_{\mu},f\right\}\dd x^{\mu}- \frac{1}{2}f_{\mu}^{\nu\lambda}(\partial_{\nu}f)A_{\lambda}\,\dd x^{\mu} \nonumber  \\
&+& \sum_{n=1}^{\infty}\frac{(-2)^n B_{2n}}{(2n)!} \frac{\lambda^{2n}  {  \mathrm{\boldsymbol{Z}} }^{n-1}}{2^n} \left( {  \mathrm{\boldsymbol{Z}} }\,\boldsymbol{\delta} {_{\mu}^{\nu}} - \boldsymbol{A}_{\boldsymbol{\alpha}}^{ {\nu}} \boldsymbol{A}_{ {\mu}} \right)(\partial_{\nu}f)\,\dd x^{\mu},
\eea
therefore the only non-zero brackets of the underlying $L_{\infty}$ algebra are given by
\bea
l_1(f) &=&  (\partial_{\mu}f) \,\dd x^{\mu}, \nonumber\\
l_2(f,A) &=&\{A_{\mu},f\} \dd x^{\mu}  - \frac{1}{2}f_{\mu}^{\nu\lambda}(\partial_{\nu}f)\,\dd x^{\mu}A_{\lambda}, \nonumber\\
l_{2n+1}(f,A,\cdots,A )  &=& \frac{ B_{2n}}{(2n)!} \cdot\lambda^{2n}  {  \mathrm{\boldsymbol{Z}} }^{n-1} \cdot\left( {  \mathrm{\boldsymbol{Z}} }\,\boldsymbol{\delta} {_{\mu}^{\nu}} - \boldsymbol{A}_{\boldsymbol{\alpha}}^{ {\nu}} \boldsymbol{A}_{ {\mu}} \right)(\partial_{\nu}f)\,\dd x^{\mu} , \quad n\in\mathbb{N}, \nonumber\\
l_2(f,g) &=& -\left\{f,g\right\},\quad\quad \forall f,g\in V_0.  \la{LinfAns}
\eea
We remind that the structure constants are given by Eq.~\eqref{ourf}, and the quantity $ {  \mathrm{\boldsymbol{Z}} }$ is defined by~\eqref{Znotations}. This result is a direct generalisation of the $L_{\infty}$ algebra, presented in the Example 6.4 of~\cite{Kupriyanov:2021cws} for the three-dimensional $\mathfrak{su}(2)$-case.

}

\section{Deformed field strength.} \label{Fsec}
According to~\cite{Kupriyanov:2020sgx, Kupriyanov:2019cug} the deformed field strength, which transforms in a covariant way under the noncommutative transformations~\eqref{gauge0},  
\be
\delta_f\mathcal{F}_{\mu\nu} = \{\mathcal{F}_{\mu\nu},f\}, \la{Ftrans}
\ee
may be searched by adapting the usual definition  of  the non-Abelian field strength to our Poisson gauge algebra. This yields  \cite{Kupriyanov:2020sgx, Kupriyanov:2019cug}:
 {\be
\mathcal{F}_{\mu\nu} = R_{\mu\nu}^{~~\,\rho\lambda} \left( 2 \gamma^{\xi}_{\rho}\, \partial_{\xi}A_{\lambda} + \{A_{\rho},A_{\lambda}\}\right), \la{Fstru}
\ee}
with the unknown $R_{\mu\nu}^{~~\xi\lambda}$  satisfying appropriate conditions.
 {It would certainly be  interesting to derive this result  from   symplectic embeddings, as we did for the gauge potential, however, such a connection is still missing. Therefore  the field strength is here obtained, as in \cite{Kupriyanov:2020sgx, Kupriyanov:2019cug}, in a more direct way.
\\
 By imposing that \eqn{Ftrans} be satisfied,  one gets an equation for the coefficient function $R_{\mu\nu}^{~~\xi\lambda}$, which we name   the second master equation,}
{\be
\gamma^{\xi}_{\lambda}\, \partial^{\lambda}_A R_{\mu\nu}^{~~\,\rho\omega} + \Theta^{\xi\lambda}\,\partial_{\lambda}R_{\mu\nu}^{~~\,\rho\omega}
+ R_{\mu\nu}^{~~\,\rho\lambda}\,\partial_{A}^{\omega}\gamma^{\xi}_{\lambda} + R_{\mu\nu}^{~~\,\lambda\omega}\,\partial_A^{\rho}\gamma^{\xi}_{\lambda} = 0. \la{master2}
\ee}
The latter exhibits the following undeformed limit:
\be
\lim_{\lambda\rightarrow 0}  R_{\mu \nu}{}^{\rho \omega}  = \frac12\left(\delta_{\mu}^{\rho}\delta_{\nu}^{\omega}-\delta_{\mu}^{\omega}\delta_{\nu}^{\rho}\right). \la{comlim2}
\ee
This requirement together with the relation~\eqref{comlim1}  {ensures} that the noncommutative field strength reduces to the commutative one in the undeformed theory:
\be
\lim_{\lambda\rightarrow 0}\mathcal{F}_{\mu\nu} = F_{\mu\nu}\equiv \partial_{\mu}A_{\nu} - \partial_{\nu}A_{\mu}. \la{Fcommlim}
\ee
%%%%%%%%%%%
%%%%%%%%%%%
A solution of Eq.~\eqref{master2}, which satisfies the condition~\eqref{comlim2}, is known for  arbitrary Poisson bivector $\Theta$ up to $\mathcal{O}(\Theta^2)$ terms~\cite{Kupriyanov:2020sgx},
 {see Appendix {\bf A}.}
Substituting our data~\eqref{Ps} in these formulae, and using the   {straightforward} identities,
\bea
f^{\rho \lambda}_{\nu}f^{\xi\omega}_{\lambda} &=& \lambda^2\big( \hat{\alpha}^{\rho\omega}\boldsymbol{\delta}^{\xi}_{\nu} - \hat{\alpha}^{\rho\xi}\boldsymbol{\delta}^{\omega}_{\nu}\big), \nonumber\\
\big(f_{\mu}^{\xi\rho}f_{\nu}^{\sigma\omega} - f_{\nu}^{\xi\rho}f_{\mu}^{\sigma\omega}\big)A_{\xi}A_{\sigma} 
&= &\lambda^2\big(\boldsymbol{\delta}_{\mu}^{\rho}\boldsymbol{\delta}_{\nu}^{\omega} - \boldsymbol{\delta}_{\nu}^{\rho}\boldsymbol{\delta}_{\mu}^{\omega}\big)\, {  \mathrm{\boldsymbol{Z}} } \nonumber\\
&+& \lambda^2\big( \boldsymbol{A}_{\boldsymbol{\alpha}}^{\omega}\boldsymbol{A}_{\mu}\boldsymbol{\delta}_{\nu}^{\rho}  
-\boldsymbol{A}_{\boldsymbol{\alpha}}^{\omega}\boldsymbol{A}_{\nu}\boldsymbol{\delta}_{\mu}^{\rho}
-\boldsymbol{A}_{\boldsymbol{\alpha}}^{\rho}\boldsymbol{A}_{\mu}\boldsymbol{\delta}_{\nu}^{\omega}
+\boldsymbol{A}_{\boldsymbol{\alpha}}^{\rho}\boldsymbol{A}_{\nu}\boldsymbol{\delta}_{\mu}^{\omega}
\big)
,
\eea
 we get:
\bea
 R_{\mu\nu}^{~~\,\rho\omega}(A) &=& \frac{1}{2}\,\Big(   
 \boldsymbol{\delta}_{\mu}^{\rho}\boldsymbol{\delta}_{\nu}^{\omega} -  \boldsymbol{\delta}_{\mu}^{\omega}\boldsymbol{\delta}_{\nu}^{\rho} \Big)\cdot 
\left(1- \frac{1}{12} \,\lambda^2  {  \mathrm{\boldsymbol{Z}} }\right) \nonumber\\
 &+& \frac{1}{2}\,\big( \delta_{\mu}^0\delta^{\rho}_0 \boldsymbol{\delta}_{\nu}^{\omega} -  \delta_{\nu}^0\delta^{\rho}_0 \boldsymbol{\delta}_{\mu}^{\omega}
 +  \delta_{\nu}^0\delta^{\omega}_0 \boldsymbol{\delta}_{\mu}^{\rho} -  \delta_{\mu}^0\delta^{\omega}_0 \boldsymbol{\delta}_{\nu}^{\rho} \big)\cdot\left(1- \frac{1}{6} \,\lambda^2  {  \mathrm{\boldsymbol{Z}} }\right)
 \nonumber\\
&+& \frac{1}{4}\,\big( f^{\xi \omega}_{\nu}\delta_{\mu}^{\rho}   -  f^{\xi \omega}_{\mu}\delta_{\nu}^{\rho}  +f^{\xi \rho}_{\mu}\delta_{\nu}^{\omega} -  f^{\xi \rho}_{\nu}\delta_{\mu}^{\omega}\big)\,A_{\xi}   \nonumber\\
&-&\frac{\lambda^2}{24} \,\big(  \boldsymbol{\delta}_{\mu}^{\rho} \boldsymbol{A}_{\boldsymbol{\alpha}}^{\omega}\boldsymbol{A}_{\nu} -  \boldsymbol{\delta}_{\nu}^{\rho} \boldsymbol{A}_{\boldsymbol{\alpha}}^{\omega}\boldsymbol{A}_{\mu}
-  \boldsymbol{\delta}_{\mu}^{\omega} \boldsymbol{A}_{\boldsymbol{\alpha}}^{\rho}\boldsymbol{A}_{\nu} +\boldsymbol{\delta}_{\nu}^{\omega} \boldsymbol{A}_{\boldsymbol{\alpha}}^{\rho}\boldsymbol{A}_{\mu}\big)  \nonumber\\
&+&\frac{\lambda^2}{12}\big( \boldsymbol{A}_{\boldsymbol{\alpha}}^{\omega}\boldsymbol{A}_{\nu} \delta_{\mu}^0\delta^{\rho}_0 -  \boldsymbol{A}_{\boldsymbol{\alpha}}^{\omega}\boldsymbol{A}_{\mu} \delta_{\nu}^0\delta^{\rho}_0
- \boldsymbol{A}_{\boldsymbol{\alpha}}^{\rho}\boldsymbol{A}_{\nu} \delta_{\mu}^0\delta^{\omega}_0 + \boldsymbol{A}_{\boldsymbol{\alpha}}^{\rho}\boldsymbol{A}_{\mu} \delta_{\nu}^0\delta^{\omega}_0 \big)  +\mathcal{O}(\lambda^3).
\eea
This formula suggests to look for the complete solution of~\eqref{master2} in the form of the following Ansatz
{\bea
 R_{\mu\nu}^{~~\,\rho\omega}(A) &=& \frac{1}{2}\,\Big(   
 \boldsymbol{\delta}_{\mu}^{\rho}\boldsymbol{\delta}_{\nu}^{\omega} -  \boldsymbol{\delta}_{\mu}^{\omega}\boldsymbol{\delta}_{\nu}^{\rho} \Big)\cdot 
  {\zeta}( \lambda \sqrt{ {  \mathrm{\boldsymbol{Z}} }}) + \frac{1}{2}\,\big( \delta_{\mu}^0\delta^{\rho}_0 \boldsymbol{\delta}_{\nu}^{\omega} -  \delta_{\nu}^0\delta^{\rho}_0 \boldsymbol{\delta}_{\mu}^{\omega}
 +  \delta_{\nu}^0\delta^{\omega}_0 \boldsymbol{\delta}_{\mu}^{\rho} -  \delta_{\mu}^0\delta^{\omega}_0 \boldsymbol{\delta}_{\nu}^{\rho} \big)\cdot\Lambda(\lambda\sqrt{ {  \mathrm{\boldsymbol{Z}} }}) \nonumber\\
&+& \frac{1}{4}\,\big( f^{\xi \omega}_{\nu}\delta_{\mu}^{\rho}   -  f^{\xi \omega}_{\mu}\delta_{\nu}^{\rho}  +f^{\xi \rho}_{\mu}\delta_{\nu}^{\omega} -  f^{\xi \rho}_{\nu}\delta_{\mu}^{\omega}\big)\,A_{\xi}  \cdot {\tilde{ {\zeta}}}( \lambda \sqrt{ {  \mathrm{\boldsymbol{Z}} }}) \nonumber\\
&+&\frac{\lambda^2}{8} \,\big(  \boldsymbol{\delta}_{\mu}^{\rho} \boldsymbol{A}_{\boldsymbol{\alpha}}^{\omega}\boldsymbol{A}_{\nu} -  \boldsymbol{\delta}_{\nu}^{\rho} \boldsymbol{A}_{\boldsymbol{\alpha}}^{\omega}\boldsymbol{A}_{\mu}
-  \boldsymbol{\delta}_{\mu}^{\omega} \boldsymbol{A}_{\boldsymbol{\alpha}}^{\rho}\boldsymbol{A}_{\nu} +\boldsymbol{\delta}_{\nu}^{\omega} \boldsymbol{A}_{\boldsymbol{\alpha}}^{\rho}\boldsymbol{A}_{\mu}\big) \cdot\phi( \lambda \sqrt{ {  \mathrm{\boldsymbol{Z}} }}) \nonumber\\
&+&\frac{\lambda^2}{8}\big( \boldsymbol{A}_{\boldsymbol{\alpha}}^{\omega}\boldsymbol{A}_{\nu} \delta_{\mu}^0\delta^{\rho}_0 -  \boldsymbol{A}_{\boldsymbol{\alpha}}^{\omega}\boldsymbol{A}_{\mu} \delta_{\nu}^0\delta^{\rho}_0
- \boldsymbol{A}_{\boldsymbol{\alpha}}^{\rho}\boldsymbol{A}_{\nu} \delta_{\mu}^0\delta^{\omega}_0 + \boldsymbol{A}_{\boldsymbol{\alpha}}^{\rho}\boldsymbol{A}_{\mu} \delta_{\nu}^0\delta^{\omega}_0 \big) \cdot \Phi( \lambda \sqrt{ {  \mathrm{\boldsymbol{Z}} }}), \la{RsolA}
\eea}
  where the form factors $ {\zeta}$, $\tilde{ {\zeta}}$, $\phi$, $\Lambda$ and $\Phi$ are unknown functions, which exhibit the following asymptotic behaviour at small $\lambda$:
\bea
 {\zeta}(\lambda\sqrt{ {  \mathrm{\boldsymbol{Z}} }}) &=& 1- \frac{1}{12} \,\lambda^2  {  \mathrm{\boldsymbol{Z}} } + \mathcal{O}(\lambda^3), \nonumber\\
\Lambda(\lambda\sqrt{ {  \mathrm{\boldsymbol{Z}} }}) &=& 1- \frac{1}{6} \,\lambda^2  {  \mathrm{\boldsymbol{Z}} } +  \mathcal{O}(\lambda^3), \nonumber\\
\tilde{ {\zeta}}(\lambda\sqrt{ {  \mathrm{\boldsymbol{Z}} }}) &=& 1 + \mathcal{O}(\lambda^2),\nonumber\\
\phi(\lambda\sqrt{ {  \mathrm{\boldsymbol{Z}} }}) &=& -\frac{1}{3} + \mathcal{O}(\lambda), \nonumber\\
\Phi(\lambda\sqrt{ {  \mathrm{\boldsymbol{Z}} }}) &=& \frac{2}{3} + \mathcal{O}(\lambda). \la{incs}
\eea
Substituting the ansatz~\eqref{RsolA} in the master equation~\eqref{master2} at $\mu = 1$, $\nu=2$, $\rho=3$, $\omega = 2$, $\xi=3$ we get, 
\be
\frac{A_1\lambda^2}{8}\left(\frac{(A_3)^2}{ {  \mathrm{\boldsymbol{Z}} }}\, u\,\left[\phi'(u) - \frac{u}{2} \,\phi(u)\,\hat{\chi}\left(\frac{u^2}{4}\right)-\frac{u}{2} \, {\zeta}(u)\,\hat{\chi}'\left(\frac{u^2}{4}\right)\right] - \left[  {\zeta}(u)\,\chi\left(\frac{u^2}{4}\right) - \phi(u)\right]\right) = 0,
\ee
where $u\equiv \lambda\sqrt{ {  \mathrm{\boldsymbol{Z}} }}$, and, we remind, the function $\hat{\chi}(v)$ is defined by Eq.~\eqref{hchidef}. This relation is satisfied for all $A_j$ iff
\bea
\phi'(u) - \frac{u}{2} \,\phi(u)\,\hat{\chi}\left(\frac{u^2}{4}\right)-\frac{u}{2} \, {\zeta}(u)\,\hat{\chi}'\left(\frac{u^2}{4}\right) &=& 0, \nonumber\\
 {\zeta}(u)\,\chi\left(\frac{u^2}{4}\right) - \phi(u) &=& 0.
\eea
The solution of this system of equations, which is compatible\footnote{Actually, one has to use just the initial condition $\phi(0) = -\frac{1}{3}$, while the asymptotic behaviour of $ {\zeta}$ can be checked a posteriori. } with the asymptotics~\eqref{incs}, is given by
\bea
 {\zeta}(u) &=& 4\left(\frac{\sin{\frac{u}{2}}}{u}\right)^2, \nonumber\\
\phi(u) &=& \frac{4\,(-2 + 2\cos{u} + u\, \sin{u})}{u^4}= \frac{2}{u}\frac{\dd  {\zeta}}{\dd u}. \la{solPart1}
\eea
In order to determine the remaining three form factors we substitute the ansatz~\eqref{RsolA} in Eq.~\eqref{master2} at $\mu = 1$, $\nu=0$, $\rho=1$, $\omega = 0$, $\xi=2$: 
\bea
0 &=& \frac{u\,\alpha}{32\,  {  \mathrm{\boldsymbol{Z}} }^{\frac{5}{2}}}\bigg(
-\sqrt{ {  \mathrm{\boldsymbol{Z}} }}\,u^2\,\alpha \left[2\,u\,\hat{\chi}'\left(\frac{u^2}{4}\right)\,\Lambda(u) + u\,\Phi(u)\,\hat{\chi}\left(\frac{u^2}{4}\right) - 4\,\Phi'(u)\right](A_1)^2A_2 \nonumber\\
&+& u^2\, {  \mathrm{\boldsymbol{Z}} }\left[u^2\,\hat{\chi}'\left(\frac{u^2}{4}\right)\,\tilde{ {\zeta}}(u) + 4\,\hat{\chi}\left(\frac{u^2}{4}\right)\,\tilde{ {\zeta}}(u) + 2\,\Phi(u)\right] A_1A_3  \nonumber\\
&-& 4\, {  \mathrm{\boldsymbol{Z}} }^{\frac{3}{2}}\left[ u\,\hat{\chi}\left(\frac{u^2}{4}\right)\,\Lambda(u) - u\,\tilde{ {\zeta}}(u) - 4\Lambda'(u)\right] A_2  \bigg), \nonumber\\
\eea
what leads us to a system of three coupled equations for three undetermined functions $\tilde{ {\zeta}}(u)$, $\Phi(u)$ and $\Lambda(u)$:
\bea
2\,u\,\hat{\chi}'\left(\frac{u^2}{4}\right)\,\Lambda(u) + u\,\Phi(u)\,\hat{\chi}\left(\frac{u^2}{4}\right) - 4\,\Phi'(u) &=& 0 \nonumber\\
u^2\,\hat{\chi}'\left(\frac{u^2}{4}\right)\,\tilde{ {\zeta}}(u) + 4\,\hat{\chi}\left(\frac{u^2}{4}\right)\,\tilde{ {\zeta}}(u) + 2\,\Phi(u) &=& 0 \nonumber\\
 u\,\hat{\chi}\left(\frac{u^2}{4}\right)\,\Lambda(u) - u\,\tilde{ {\zeta}}(u) - 4\Lambda'(u) &=& 0,\quad u\equiv \lambda\sqrt{ {  \mathrm{\boldsymbol{Z}} }}.
\eea
Resolving these equations, and imposing the conditions~\eqref{incs}, we obtain:
\bea
\tilde{ {\zeta}}(u) &=& 4\left(\frac{\sin{\frac{u}{2}}}{u}\right)^2 =  {\zeta}(u), \nonumber\\
\Lambda(u) &=& \frac{\sin{u}}{u}, \nonumber\\
\Phi(u) &=& \frac{4\,(u- \sin{u})}{u^3}. \la{solPart2}
\eea
Summarising Eq.~\eqref{RsolA}, Eq.~\eqref{solPart1} and  Eq.~\eqref{solPart2} we arrive at
 {\bea
 R_{\mu\nu}^{~~\,\rho\omega}(A) &=& \frac{1}{2}\,\Big(   
 \boldsymbol{\delta}_{\mu}^{\rho}\boldsymbol{\delta}_{\nu}^{\omega} -  \boldsymbol{\delta}_{\mu}^{\omega}\boldsymbol{\delta}_{\nu}^{\rho} \Big)\cdot 
  {\zeta}( \lambda \sqrt{ {  \mathrm{\boldsymbol{Z}} }}) + \frac{1}{2}\,\big( \delta_{\mu}^0\delta^{\rho}_0 \boldsymbol{\delta}_{\nu}^{\omega} -  \delta_{\nu}^0\delta^{\rho}_0 \boldsymbol{\delta}_{\mu}^{\omega}
 +  \delta_{\nu}^0\delta^{\omega}_0 \boldsymbol{\delta}_{\mu}^{\rho} -  \delta_{\mu}^0\delta^{\omega}_0 \boldsymbol{\delta}_{\nu}^{\rho} \big)\cdot\Lambda(\lambda\sqrt{ {  \mathrm{\boldsymbol{Z}} }}) \nonumber\\
&+& \frac{1}{4}\,\big( f^{\xi \omega}_{\nu}\delta_{\mu}^{\rho}   -  f^{\xi \omega}_{\mu}\delta_{\nu}^{\rho}  +f^{\xi \rho}_{\mu}\delta_{\nu}^{\omega} -  f^{\xi \rho}_{\nu}\delta_{\mu}^{\omega}\big)\,A_{\xi}  \cdot { {\zeta}}( \lambda \sqrt{ {  \mathrm{\boldsymbol{Z}} }}) \nonumber\\
&+&\frac{\lambda^2}{8} \,\big(  \boldsymbol{\delta}_{\mu}^{\rho} \boldsymbol{A}_{\boldsymbol{\alpha}}^{\omega}\boldsymbol{A}_{\nu} -  \boldsymbol{\delta}_{\nu}^{\rho} \boldsymbol{A}_{\boldsymbol{\alpha}}^{\omega}\boldsymbol{A}_{\mu}
-  \boldsymbol{\delta}_{\mu}^{\omega} \boldsymbol{A}_{\boldsymbol{\alpha}}^{\rho}\boldsymbol{A}_{\nu} +\boldsymbol{\delta}_{\nu}^{\omega} \boldsymbol{A}_{\boldsymbol{\alpha}}^{\rho}\boldsymbol{A}_{\mu}\big) \cdot\phi( \lambda \sqrt{ {  \mathrm{\boldsymbol{Z}} }}) \nonumber\\
&+&\frac{\lambda^2}{8}\big( \boldsymbol{A}_{\boldsymbol{\alpha}}^{\omega}\boldsymbol{A}_{\nu} \delta_{\mu}^0\delta^{\rho}_0 -  \boldsymbol{A}_{\boldsymbol{\alpha}}^{\omega}\boldsymbol{A}_{\mu} \delta_{\nu}^0\delta^{\rho}_0
- \boldsymbol{A}_{\boldsymbol{\alpha}}^{\rho}\boldsymbol{A}_{\nu} \delta_{\mu}^0\delta^{\omega}_0 + \boldsymbol{A}_{\boldsymbol{\alpha}}^{\rho}\boldsymbol{A}_{\mu} \delta_{\nu}^0\delta^{\omega}_0 \big) \cdot \Phi( \lambda \sqrt{ {  \mathrm{\boldsymbol{Z}} }}), \la{Rsol}
\eea}
with
 {\bea
 {\zeta}(u) &=& 4\left(\frac{\sin{\frac{u}{2}}}{u}\right)^2, \nonumber\\
\Lambda(u) &=& \frac{\sin{u}}{u}, \nonumber\\
\phi(u) &=& \frac{2}{u}\frac{\dd \lambda}{\dd u}, \nonumber\\
\Phi(u) &=& \frac{4\,(u- \sin{u})}{u^3}. \la{formfactors}
\eea}
One can check by  direct substitution that our solution is valid for all other combinations of the indexes $\mu$, $\nu$, $\rho$, $\omega$ and $\xi$.

At $\alpha=1$ the three-dimensional restriction, $R_{ab}^{~~\,cd}$, of \eqn{Rsol} coincides\footnote{We use slightly different parametrisation of the form factors.} with the known three-dimensional solution for the $\mathfrak{su}(2)$ case~\cite{Kupriyanov:2020sgx}.  {It is remarkable that the presence of the fourth (commutative) coordinate $x^0$ generalises the mentioned three-dimensional result in a quite nontrivial way, introducing new contributions of the form factors $\Lambda$ and $\Phi$.

The deformed field strength $\mathcal{F}$, defined by Eq. \eqn{Fstru}, allows for a natural definiton of the classical action functional, which remains invariant upon the deformed noncommutative gauge transformations~\eqref{gauge0}, and which reproduces correctly the classical limit. Indeed, by defining
\be
S[A] := \int_{\mathbb{R}^4} \dd^4 x \, \mathcal{L}, \quad\quad \mathcal{L} := -\frac{1}{4} \mathcal{F}_{\mu\nu}\mathcal{F}_{\rho\xi} \,\eta^{\mu\rho}\eta^{\nu\xi}, \la{Sdef}
\ee
with
\be
\eta = \mathrm{diag}\,(+1,-1,-1,-1),
\ee
we can check  that the classical limit is 
\be
\lim_{\lambda\rightarrow 0} S[A] = \int_{\mathbb{R}^4} \dd^4 x \, \Big( - \frac{1}{4}F_{\mu\nu} F_{\rho\xi}\,\eta^{\mu\rho}\eta^{\nu\xi}\Big),
\ee
thanks to the property~\eqref{Fcommlim} of the deformed field strength.
Moreover, since the deformed field strength $\mathcal{F}$ transforms in a covariant way~ (Eq. \eqref{Ftrans}), the deformed Lagrangian density,  being quadratic in $\mathcal{F}$, transforms in a covariant way as well. We have indeed
\bea
\delta_{f}\mathcal{L} 
&\stackrel{\text{ {def}}}{ {=}}&   {-\frac{1}{4}\lim_{\varepsilon\to 0}\frac{\partial}{\partial \varepsilon} \big(\mathcal{F}_{\mu\nu} + \varepsilon\,(\delta_f\mathcal{F}_{\mu\nu}) \big)\,\big(\mathcal{F}_{\rho\xi} + \varepsilon\, (\delta_f\mathcal{F}_{\rho\xi})\big) \,\eta^{\mu\rho}\eta^{\nu\xi} }\nonumber\\
  & {=}& {-\frac{1}{4}\Big(\mathcal{F}_{\mu\nu}\,\big(\delta_f\mathcal{F}_{\rho\xi}\big) + \big(\delta_f\mathcal{F}_{\mu\nu}\big) \mathcal{F}_{\rho\xi}\Big)\,\eta^{\mu\rho}\eta^{\nu\xi} 
   }
   \nonumber\\
  & {=}&   {-\frac{1}{4}\Big(\mathcal{F}_{\mu\nu}\,\big\{\mathcal{F}_{\rho\xi},f\big\} + \big\{\mathcal{F}_{\mu\nu},f\big\} \mathcal{F}_{\rho\xi}\Big)\,\eta^{\mu\rho}\eta^{\nu\xi} 
  =-\frac{1}{4}\,\big\{ \mathcal{F}_{\mu\nu}\,\mathcal{F}_{\rho\xi},f\big\} \,\eta^{\mu\rho}\eta^{\nu\xi}
  } \nonumber\\
&=& \{\mathcal{L},f\} , \la{calculuS}
\eea
 {where we have first used the  standard definition of  first variation of $\mathcal{L}$ upon the variation of $\mathcal{F}$; then
we have substituted the explicit expression~\eqref{Ftrans} for $\delta_f\mathcal{F}$, and took into account the derivation property of the Poisson bracket.
By using 
 Eq.~\eqref{PoissBrack} we thus get
 \be
 \delta_{f}\mathcal{L}
= \partial_{\mu}(\mathcal{L} \,\,\partial_{\nu}f\,\, \Theta^{\mu\nu}) - \mathcal{L}\,\,\partial_{\nu}f\,\, \partial_{\mu}\Theta^{\mu\nu}.\label{calculuS2}
\ee
\noindent{\bf Remark.} In order to avoid confusions, we comment on the usage of partial derivatives in this paper. On the one hand, in the master equations~\eqref{master1} and \eqref{master2}, in the terms $ \Theta^{\nu\mu} \partial_{\mu} \gamma^{\xi}_{\lambda} 
- \Theta^{\xi\mu} \partial_{\mu} \gamma^{\nu}_{\lambda} $ and $ \Theta^{\xi\lambda}\,\partial_{\lambda}R_{\mu\nu}^{~~\,\rho\omega}$, the partial derivatives  act on the \emph{explicit} dependence on $x$ only, whilst $A(x)$ is considered as an independent variable.  On the other hand, in all other places of this article, (e.g. in the definition 
of the Poisson bracket~\eqref{PoissBrack}), the partial derivatives act on \emph{all}  $x$-dependent objects. In particular, in the last line of Eq.~\eqref{calculuS2} 
 the partial derivative acts on $x$, which is present in $\mathcal{L}$ not just explicitly \footnote{ {i.e. via the Poisson bivector $\Theta(x)$, which enters in the definition~\eqref{Fstru} of $\mathcal{F}$ through the Poisson bracket. }}, but also via $A(x)$ and its first derivatives as well. This  justifies  the Leibnitz rule in the last step of~\eqref{calculuS2}.
}

Finally, noticing that the Poisson bivector~\eqref{Ps} satisfies the identity
\be
\partial_{\mu}\Theta^{\mu\nu} = 0,
\ee
we see that the variation $\delta_{f}\mathcal{L}$ is a total derivative, therefore the action~\eqref{Sdef} is gauge invariant:
\be
\delta_f S[A] = 0.
\ee
}
 {\section{Summary and outlook.} In this article we constructed a family of four-dimensional noncommutative deformations of the  $U(1)$ gauge theory, implementing a class of noncommutative spaces~\eqref{Ps} in the general framework of~\cite{Kupriyanov:2020sgx}. This class includes the angular (or $\lambda$-Minkowski), the $\mathfrak{su}(2)$ and the $\mathfrak{su}(1,1)$ cases at $\alpha = 0$, $\alpha=+1$ and $\alpha=-1$ respectively.  {We worked within the semi-classical approximation, so our noncommutative gauge theories are actually  Poisson gauge theories.}

The first %key 
result is the definition~\eqref{gauge0} of  deformed gauge transformations, where the matrix  $\gamma$ is given  by Eq.~\eqref{GammAsol}. These transformations close the noncommutative
 algebra~\eqref{gaugealgebra}. {We also {discussed} the interpretation of  {the master equation Eq.~\eqref{master1}, which  we used to construct the deformed gauge transformations}, as a Jacobi identity for  symplectic embeddings \cite{Kupriyanov:2021cws, Kupriyanov:2021aet}.}

The second %key 
result is an explicit $L_{\infty}$ structure~\eqref{LinfAns}, which corresponds to our deformed  {noncommutative} gauge transformations in sense of the $L_{\infty}$ bootstrap. 

The third %key
 result is  an expression for the deformed field strength, Eq.~\eqref{Fstru}, where the quantity $R$ is given by Eq.~\eqref{Rsol}. 
This deformed field strength transforms in a covariant way upon the deformed  {noncommutative} gauge transformations, thereby allowing for a definition of  the gauge-invariant classical action~\eqref{Sdef}. We stress  that the presence of the fourth (commutative) coordinate $x^0$ brings nontrivial contributions to the deformed strength (via $R$), which do  {not} look like a simple and intuitive addition to the corresponding three-dimensional result.  {
In particular, the components $\mathcal{F}_{0j}$ exhibit a highly nonlinear dependence on the three-dimensional components of $A$. This behaviour  is different from   the one of the matrix $\gamma$, where the  four-dimensionality does not change the corresponding three-dimensional result that much, since $\gamma^{0\rho} = \delta^{0\rho}$. 
Let us illustrate the nontriviality on a simple example where the   gauge potential does not depend on spatial coordinates $x^{j}$. In this situation one may expect that the noncommutativity, being essentially three-dimensional, does not affect the field strength, so $\mathcal{F}_{0j} = \partial_0 A_j$. Our analysis, instead, yields 
\bea
\mathcal{F}_{0j} &=& 2\, R_{0 j}^{~~\,0 k}\,\partial_0A_k  \nonumber\\ 
&=& \Lambda(\lambda \sqrt{\mathrm{\boldsymbol{Z}}}) \,\partial_0A_j  + \frac{1}{2}\,\zeta(\lambda \sqrt{\mathrm{\boldsymbol{Z}}})\, f^{rk}_j A_r\,\partial_0A_k 
+ \frac{\lambda^2}{8} \, \Phi(\lambda \sqrt{\mathrm{\boldsymbol{Z}}}) \,\boldsymbol{A}_{\boldsymbol{\alpha}}^{k}A_{j}\,\partial_0A_k,
\eea
which disproves  the naive expectation. \emph{The nonlinearity which derives from  spatial noncommutativity manifests itself in the spatially-homogeneous  situation as well, as far as the time-dependence is concerned.} 
One can also check that, substituting an $x^{j}$-independent gauge potential $A$ in the equations of motion derived   from the classical action~\eqref{Sdef}, one gets a nonlinear dynamics, which governs the time-dependence.  At the best of our knowledge nothing similar takes place in  noncommutative gauge theories which are based on  more conventional  approaches.
% than the one  considered here, which is based on  
%~\cite{Kupriyanov:2020sgx}.%, which connects the spatial noncommutativity directly to the (deformed) gauge algebra. 
}

The present research can be continued in various directions. On  one hand, one may study various physical consequences of  noncommutativity such as the existence of Gribov copies,  which has already been established for the noncommutative QED with  Moyal type noncommutativity~\cite{Canfora:2015nsa, Kurkov:2017tyn, Holanda:2021lhk}. On the other hand, one may focus on purely mathematical structures, which stand behind. In particular, one may wonder whether the field strength $\mathcal{F}$, obtained in this paper, is compatible with the $L_{\infty}$ bootstrap procedure, related to the \emph{extended} $L_{\infty}$ algebra~\cite{Blumenhagen:2018kwq}, that contains one more nonempty subspace $V_{-2}$ of the objects which transform in a covariant way upon the deformed gauge transformations. In particular, one has to check, whether
\be
\mathcal{F}_{\mu\nu} \,dx^{\mu}\wedge dx^{\nu} = \sum_{n=1}^{\infty} \frac{1}{n!}(-1)^{\frac{n(n-1)}{2}} l_n(A,\cdots,A), 
\ee
where the first bracket corresponds to the undeformed field strength~\eqref{Fcommlim},
\be
l_1(A) = F_{\mu\nu} \,dx^{\mu}\wedge dx^{\nu},
\ee
and all the brackets together fulfil the $L_{\infty}$ relations.

 {Finally, an interesting problem which we would like to investigate in the coming future is to understand  if   it is possible to derive the deformed field strength proposed in this paper within the symplectic embedding approach, and to clarify  its geometric nature. }

%QQQQQ

}
\section*{A. General solution for $R_{{\mu}{\nu}}{}^{{\rho}{\omega}}(x,A) $ up to $\mathcal{O}\big(\Theta^3\big)$ corrections.}
\be
R_{{\mu}{\nu}}{}^{{\rho}{\omega}}(x,A) = R_{{\mu}{\nu}}^{(0)\,{\rho}{\omega}}(x,A) + R_{{\mu}{\nu}}^{(1)\,{\rho}{\omega}}(x,A) + R_{{\mu}{\nu}}^{(2)\,{\rho}{\omega}}(x,A) + \mathcal{O}\big(\Theta^3\big),
\ee
where
\bea
R_{{\mu}{\nu}}^{(0)\,{\rho}{\omega}}(x,A) &=& \frac12\left(\delta_{\mu}^{{\rho}}\delta_{\nu}^{{\omega}}-\delta_{\mu}^{{\omega}}\delta_{\nu}^{{\rho}}\right), \nonumber\\
R_{{\mu}{\nu}}^{(1)\,{\rho}{\omega}}(x,A) &=&
\frac14\left(\delta^{\rho}_{\mu}\,\partial_{\nu}\Theta^{{\xi}{\omega}}-\delta^{\omega}_{\mu}\,\partial_{\nu}\Theta^{{\xi}{\rho}}-\delta^{\rho}_{\nu}\,\partial_{\mu}\Theta^{{\xi}{\omega}}+\delta^{\omega}_{\nu}\,\partial_{\mu}\Theta^{{\xi}{\rho}}\right)A_{\xi}, \nonumber\\
R_{{\mu}{\nu}}^{(2)\,{\rho}{\omega}}(x,A) &=&\left(\frac{1}{12}\delta^{\rho}_{\mu}\,\Theta^{{\sigma}{\phi}}\,\partial_{\nu}\partial_{\phi}\Theta^{{\xi}{\omega}}-\frac{1}{12}\delta^{\omega}_{\mu}\,\Theta^{{\sigma}{\phi}}\,\partial_{\nu}\partial_{\phi}\Theta^{{\xi}{\rho}} \right.\nonumber\\
&-& \frac{1}{12}\delta^{\rho}_{\nu}\,\Theta^{{\sigma}{\phi}}\,\partial_{\mu}\partial_{\phi}\Theta^{{\xi}{\omega}}+\frac{1}{12}\delta^{\omega}_{\nu}\,\Theta^{{\sigma}{\phi}}\,\partial_{\mu}\partial_{\phi}\Theta^{{\xi}{\rho}} \nonumber\\
&+&\frac{1}{12}\delta^{\rho}_{\mu}\,\partial_{\nu}\Theta^{{\sigma}{\phi}}\,\partial_{\phi}\Theta^{{\xi}{\omega}}-\frac{1}{12}\delta^{\omega}_{\mu}\,\partial_{\nu}\Theta^{{\sigma}{\phi}}\,\partial_{\phi}\Theta^{{\xi}{\rho}}  \nonumber\\
&-& \frac{1}{12}\delta^{\rho}_{\nu}\,\partial_{\mu}\Theta^{{\sigma}{\phi}}\,\partial_{\phi}\Theta^{{\xi}{\omega}}+\frac{1}{12}\delta^{\omega}_{\nu}\,\partial_{\mu}\Theta^{{\sigma}{\phi}}\,\partial_{\phi}\Theta^{{\xi}{\rho}}\nonumber\\
&+&\left.\frac{1}{8}\partial_{\mu}\Theta^{{\xi}{\rho}}\,\partial_{\nu}\Theta^{{\sigma}{\omega}}-\frac{1}{8}\partial_{\mu}\Theta^{{\sigma}{\omega}}\,\partial_{\nu}\Theta^{{\xi}{\rho}}\right)\,A_{\xi}A_{\sigma}.
\eea
 {\noindent{\bf Acknowledgments.} The authors are grateful to Vlad Kupriyanov for  discussions and useful suggestions. }

\end{document}